\documentclass[runningheads]{llncs}
\usepackage{graphicx}

\begin{document}
\title{Why Fair Automated Hiring Systems Breach EU Non-Discrimination Law\thanks{The research presented in this paper has received funding from the European Union’s funded project LeADS under Grant Agreement no. 956562. A special thanks to Onntje Hinrichs and Giovanni Comande' for their guidance.}}

\author{Robert Lee Poe\inst{1}}
%

\institute{Sant'Anna School of Advanced Studies, Pisa 56127, Italy\\
\email{robertlpoe@icloud.com}}

\maketitle              

\begin{abstract}
Employment selection processes that use automated hiring systems based on machine learning are becoming increasingly commonplace. Meanwhile, concerns about algorithmic direct and indirect discrimination that result from such systems are front-and-center, and the technical solutions provided by the research community often systematically deviate from the principle of equal treatment to combat disparate or adverse impacts on groups based on protected attributes. Those technical solutions are now being used in commercially available automated hiring systems, potentially engaging in real-world discrimination. Algorithmic fairness and algorithmic non-discrimination are not the same. This article examines a conflict between the two: whether such hiring systems are compliant with EU non-discrimination law.

\end{abstract}

\newpage\section{Introduction}\label{intro}

In employment decisions, equality is sought in opportunity or outcome. Equality of opportunity can be defined as formal or substantive. Formal equality of opportunity requires that applicants be assessed according to their qualifications, that those qualifications be appropriate,\footnote{Where appropriateness is defined in relation to moral relevance \cite{arif_khan_towards_2022} or to the lawfulness of desiderata in accordance with indirect discrimination doctrine \cite{de_mol_novel_2011}.} and that the most qualified applicant receives the position \cite{arneson_equality_2015}. Selection processes that enforce formal equality of opportunity result in inequalities in outcome between groups when the individuals of a given group are, on a whole, less qualified than another in a given field.\footnote{\emph{See} e.g. Plato's \emph{Laws} discussing the notion of equality, ``[W]hen equality is given to unequal things, the resultant will be unequal . . .'' \cite[757a]{plato_plato_1914}; \emph{see} also Hayek on Equality, Value and Merit, ``From the fact that people are very different it follows that, if we treat them equally, the result must be inequality in their actual position . . .'' \cite[p. 87]{hayek_constitution_1976}.} While substantive equality of opportunity requires all that formal equality of opportunity insists upon during a selection process, it is first and foremost an effort to ensure that each individual in society, regardless of their group membership, has the same opportunities to gain the prerequisite qualifications for positions so that differences between groups are minimal or nonexistent.\footnote{\emph{See}, for instantiations of substantive equality of opportunity, e.g. Recommendation 84/635/EEC on the promotion of positive action for women (OJ 1984 L 331, p.34), Charter of Fundamental Rights of the European Union (7 December 2000), Art. 23, and Directive 2000/43/EC of 29 June 2000 implementing the principle of equal treatment between persons irrespective of racial or ethnic origin, Art.5. } Equality of outcome, also known as \emph{equity}, requires group equality or similarity in results, irrespective of the differences individuals of those groups may have in terms of qualifications for a given position\textemdash
generally for the purpose of providing a shortcut from opportunity to representation when substantive equality has not yet been fully realized.\footnote{Case 450/93 \emph{Kalanke v Bremen} [1995] ECR I-3051, § 23.}

In the European Union, \emph{positive action} is an umbrella term used to describe soft measures like the voluntary pruning of facially neutral employment criteria that may lead to disparate impacts, mainstreaming initiatives, accommodations, the use of impact assessments, and outreach programs \cite[p. 11]{commission_beyond_2007} for achieving substantive equality of opportunity for members of groups that deal with the consequences of past or present discrimination or disadvantage so that they may compete on an equal footing with others; whereas \emph{positive discrimination} is a term used to describe strong measures for achieving equality of outcome through preferential treatment when, in a given field of employment, members of the discriminated against or disadvantaged group are not yet, on the whole, equally qualified.\footnote{\emph{See} \cite[p. 442-462]{van_caeneghem_legal_2019} for a detailed explanation of the difference between positive action and positive discrimination in EU non-discrimination law; \emph{see also} \cite[p. 6]{bell_putting_2007}. \emph{But see} e.g. \cite[p.34]{hacker_teaching_2018} and \cite[p. 752-4]{wachter_bias_2021} for the conflation of the two distinct concepts in the fair machine learning literature.} Positive discrimination in employment decisions is controversial, and the practice has been repeatedly restrained by the Court of Justice of the European Union (CJEU) whenever employment selection processes move from the goal of ensuring formal and substantive equality of opportunity into the pursuit of equal outcomes (See Section \ref{PD}). The CJEU case-law pertaining to positive discrimination in employment has been settled for nearly two decades, and legal scholars have repeatedly concluded that the CJEU ``systematically rejects'' selection processes that turn towards equality of outcome \cite[p. 462, 587]{van_caeneghem_legal_2019}\cite[p. 6]{bell_putting_2007}\cite[p. 356-9]{ocinneide_positive_2006}.

Automated hiring systems based on machine learning are becoming increasingly commonplace, concerns about algorithmic indirect discrimination in employment decisions are front-and-center, and the technical solutions provided by the research community often systematically deviate from the principle of equal
treatment to combat disparate impacts.\footnote{Such practices are usually referred to as `fairness-aware machine learning' \cite{dwork_fairness_2012}.} The study of algorithmic discrimination and the corresponding solutions fall under a multi-disciplinary domain known as \emph{fair machine learning}. There, legal scholarship on algorithmic discrimination has predominately focused on analyzing the training data of automated systems for features that, if used, may constitute direct or indirect discrimination and the corresponding decisions of those systems for disparate or adverse outcomes \cite{wachter_why_2021}\cite{xenidis_eu_2019}. 

Indirect  or ``covert" discrimination is understood in contrast with direct or ``overt" discrimination \cite[p. 394]{wouters_equality_2021} and is fundamentally aimed at achieving substantive equality of opportunity \cite[p. 69-72]{de_vos_european_2020-1}. Indirect discrimination takes place when a neutral provision, criterion, or practice results in a disparate impact on a protected group, ``unless that provision, criterion or practice is objectively justified by a legitimate aim and the means of achieving that aim are appropriate and necessary."\footnote{\emph{See} Art. 2(b) of Directive 2004/113/EC; \emph{see also} Art. 2(1)(b) of the Race Equality Directive; Art. 2(1)(b) of Directive 2006/54/EC.} Thus, using a hiring criteria that causes a disparate impact is not automatically discriminatory. Instead, such criterion are only discriminatory if the principle of proportionality is violated. The proportionality test, therefore, ``opens the path for the legality of using a factor that correlates with economically or otherwise favorable traits even though the choice of that factor also leads to the unfavorable treatment of a protected group" \cite[p. 17]{hacker_teaching_2018}. 

Whether a legal analysis determining the lawfulness of using features that result in a disparate impact based on a protected attribute is performed in practice by designers of automated hiring systems is difficult to know and beyond the scope of this article. It is clear, however, that designers of these systems are aware that criteria which cause a disparate impact based on a sensitive attribute can potentially be deemed discriminatory \cite{feldman_certifying_2015}\cite{barocas_big_2016}. Their solution: require equality or similarity in employment outcomes. For example, authors in \cite{raghavan_mitigating_2020} investigated automated hiring systems and found that a number of the commercially available systems for pre-selection either remove or curate the training data that produce a disparate or adverse impact or modify the objective function of the learning algorithm to achieve the same result, often in accordance with the \emph{disparate impact metric} \cite[p.471-3]{raghavan_mitigating_2020}.

It is widely recognized that automated hiring systems must not discriminate. Often fair machine learning and the tool-set it provides is seen as the answer to creating non-discriminatory automated hiring systems. However, the fact is that the implementation of a number of fair machine learning metrics and techniques ensure a discriminatory effect, albeit a positive one. An automated hiring system that, in an effort to achieve ``fairness,'' eliminates differences between groups based on a sensitive attribute while disregarding the base-rate differences between those groups, positively discriminates against applicants belonging to the ``over-represented'' or ``priveledged'' group, and the severity of that discrimination is dependent on the strength of the correlation between the sensitive attribute and the target variable in the original, unmodified sample (See Section \ref{FML}). While ``fairness'' and non-discrimination are often used synonymously or at least in the same breath, the importance of ``accuracy'' and the estimation and preservation of model generalizability should not be ignored when determining the legality of such systems with non-discrimination law.

\section{Positive Discrimination, Employment, and the CJEU}\label{PD}

There are a number of equality directives in EU law.\footnote{\emph{See} Directive 2000/43/EC of 29 June 2000 implementing the principle of equal treatment between persons irrespective of racial or ethnic origin; Directive 2000/78/EC of 27 November 2000 establishing a general framework for equal treatment in employment and occupation; Directive 2006/54/ EC of 5 July 2006 on the implementation of the principle of equal opportunities and equal treatment of men and women in matters of employment and occupation (formerly the Directive 76/207 which will be specifically analyzed in the case-law below).} 

Each directive is an embodiment of the principle of equal treatment. Equal treatment means that there shall be no discrimination \emph{whatsoever}, either directly or indirectly, based on the protected attribute laid out in a given directive. However, each equality directive moves from formal to substantive equality of opportunity by allowing Member States to adopt \emph{special measures} to prevent or compensate for disadvantages linked to the protected attribute. Thus, while the exception to the individual right of equal treatment must be interpreted strictly,\footnote{\emph{Kalanke}, supra note 4, at § 21 citing Case 222/84 Johnston v. Chief Constable of the Royal Ulster Constabulary [1986] ECR 1651, § 36.} measures which take advantage of the derogation, while discriminatory in appearance, ``are in fact intended to eliminate or reduce actual instances of inequality which may exist in the reality of social life."\footnote{\emph{Id.} at § 18-19. \emph{See also} Case 409/95 \emph{Marschall v Land Nordrhein-Westfalen} [1997] ECR I-6363, § 26-27; Case 158/97 \emph{Badeck v Hessischer Ministerpresident} [2000] ECR I-1875, § 19; and Case 476/99 \emph{Lommers v Minister van Landbouw, Natuurbeheer en Visserij} [2002] ECR I-02891, § 32.}

For instance, in \emph{Badeck}, the Court drew a distinction between training opportunities and employment opportunities. The Court applied a substantive view of equality of opportunity, allowing for the reservation of training slots for the underrepresented sex, because training opportunities are precisely where the underrepresented sex can receive the qualifications required for employment positions.\footnote{\emph{Badeck}, supra note 10 at § 52-55.} The \emph{Badeck} Court also allowed, under the derogation, a national rule that guarantees that qualified women, who satisfy all the conditions required for the position, are called to interview in sectors in which they are under-represented, because such provisions do not ``imply an attempt to achieve a final result."\footnote{\emph{Id.} at § 60-63 citing Opinion of Advocate General Saggio delivered on 10 June 1999, § 41.} In other words, such provisions do not sacrifice equal treatment for the sake of equal outcomes. 

Before moving forward, two caveats should be noted. First, the CJEU has only analyzed gender-based provisions in the employment context through the exception under Art. 2(4) of Directive 76/207 (now Directive 2006/54/EC) and Art. 141.4 of the TEC (now Article 157.4 of the TFEU).\footnote{Case 407/98 \emph{Abrahamsson and Andersson v Fogelqvist} [2000] ECR I-5539, § 62.} To what extent the exceptions of other equality directives will be treated similarly is a matter of academic debate \cite[p.355]{ocinneide_positive_2006}. Second, while the CJEU has held that direct horizontal effect can be found in the relationship between equality directives and the Charter of Fundamental Rights, that legal issue will not be discussed here.\footnote{\emph{See} \cite{de_mol_novel_2011} for more information on the topic.} As the reader moves through the following case-law, bear in mind the distinction between soft positive action measures implemented to provide substantive equality of opportunity, like the ones described above, with strong positive discrimination measures implemented to provide equality of outcome, such as the ones under review in the following cases.

\subsection{Kalanke v. Bremen}
In \emph{Kalanke}, two candidates were shortlisted for the position of Section Manager at the Bremen Parks Department in Germany. Mr. Kalanke, one of the two candidates, had a diploma in horticulture and landscape gardening, had worked for the Parks Department since 1973, and had been acting as the permanent assistant to the previous Section Manager before the position was vacated; Ms. Glibmann, the other candidate, also had a diploma in landscape gardening, granted in 1983, and had worked in the Parks Department as a horticultural employee since 1975. The Parks Department management put forward Mr. Kalanke for the position, but the Staff Committee refused its consent to his promotion.\footnote{\emph{Kalanke}, supra note 4, at § 4 and 5.}

The Staff Committee refused its consent for the promotion of Mr. Kalanke in accordance with the Bremen Law on Equal treatment for Men and Women in the Public Services (LGG) passed in 1990, which stated that women who have the same qualifications as men, applying for the same post, are to be given priority where women are underrepresented in the sector. Mr. Kalanke was successful in arbitration, but the Staff Committee appealed to the conciliation board where the two candidates were found to be equally qualified and priority was to be given to Ms. Glibmann. The case made its way through the labor courts, and eventually the Bundesarbeitsgericht sought a preliminary ruling from the CJEU clarifying the scope of the exception under Article 2 (4) of the Directive from the principle of equal treatment.\footnote{\emph{Id.} at § 3, 5, and 6-11.}

The Court began by stating that the purpose of Directive 76/207 is to put into effect the principle of equal treatment for men and women regarding access to employment and promotion within Member States, and that the principle of equal treatment means that there shall be no discrimination whatsoever, either directly or indirectly, on the grounds of sex. The exception under Article 2 (4) of the Directive 76/207 permits national measures which, although discriminatory in appearance, are intended to eliminate or reduce actual instances of inequality and consequently give a specific advantage to women with a view to improving their ability to compete on the labor market and to pursue a career on an equal footing with men.\footnote{\emph{Id.} at § 17-19 citing Case 312/86 \emph{Commission v. France} [1988] ECR 6315, § 15.} Since Article 2(4) is a derogation from an individual right, the Court determined that the exception must be strictly interpreted.\footnote{\emph{Id.} at § 21 citing Case 222/84 \emph{Johnston v. Chief Constable of the Royal Ulster Constabulary} [1986] ECR 1651, § 36.}

The Court found that national rules that guarantee women \emph{absolute} and \emph{unconditional} priority go beyond promoting equal opportunities and overstep the limits of the exception. The Court reasoned that such measures take a shortcut from ensuring substantive equality in fact to mere equality in outcome: 
\begin{quote}
    ``Furthermore, in so far as it seeks to achieve equal representation of men and women in all grades and levels within a department, such a system substitutes for equality of opportunity as envisaged in Article 2(4) the result which is only to be arrived at by providing such equality of opportunity.''
\end{quote}
Thus, the Court ruled that Art. 2(1) and (4) of the Directive 76/207 precludes national rules which \emph{automatically} give priority to women in sectors where they are underrepresented.\footnote{\emph{Id.} at § 22-24.}

\subsection{Marschall v. Land Nordrhein-Westfalen}

In 1994, a teacher named Mr. Marschall applied for a promotion to an open position at a German comprehensive school. In response, Mr. Marschall was informed that, in accordance with the civil service law of the Land, a female candidate of equal suitability, competence and professional performance was to be appointed to the position because there were fewer women than men in that particular grade post in the career bracket. Mr. Marschall brought legal action. The Administrative Court of Gelsenkirchen found that the outcome of the case was dependent on the compatibility of the Land's provision with Art. 2(1) and (4) of Directive 76/207 and so a preliminary ruling was sought from the CJEU.\footnote{\emph{Marschall}, supra note 10 at § 6-10.}

The Court began by distinguishing the case from \emph{Kalanke}. Unlike in \emph{Kalanke}, the provision in question contained a `savings clause' that stated that where an individual male candidate had qualifications that might tilt the balance in his favor, a female candidate would not be given priority. After citing the third recital in the preamble to Recommendation 84/635/EEC on the promotion of positive action for women, which highlights the need for positive action to counteract prejudices that arise in the employment context due to social attitudes, behaviors, and structures, the Court agreed with the Land and other governments that, even when candidates of the opposite sex are equally qualified, male candidates tend to be promoted in preference to female candidates because of a multitude of stereotypes. Thus, ``. . . the mere fact that a male candidate and a female candidate are equally qualified does not mean that they have the same chances.''\footnote{\emph{Id.} at § 28-30.}

The Court reasoned that a national rule may be lawful under Article 2 (4) if, in each individual case:
\begin{quote}
    ``it provides for male candidates who are equally as qualified as the female candidates a guarantee that the candidatures will be the subject of an \emph{objective assessment} which will take account of all criteria specific to the individual candidates and will override the priority accorded to female candidates where one or more of those criteria tilts the balance in favour of the male candidate. In this respect, however, it should be remembered that those criteria must not be such as to discriminate against female candidates'' [emphasis added].
\end{quote}
Thus, the Court ruled that a national rule which, conditional on a guarantee that the candidatures will be subject to an objective assessment on an individual basis and where that objective assessment tilts in the favor of a male candidate the priority will be overridden, provides a priority to female candidates who are equally qualified, with the purpose of counteracting prejudiced tie-breaking, is compatible with Art. 2(1) and (4) of Directive 76/207.\footnote{\emph{Id.} at § 33-35.}

\subsection{Abrahamsson and Anderson v. Fogelqvist}

In 1996, eight candidates applied for a professorship at the University of Göteborg, including Ms. Abrahamsson, Ms. Destouni, Ms. Fogelqvist, and Mr. Anderson. The selection board voted twice: (1) in relation to the scientific qualifications of all candidates, Mr. Anderson received five votes and Ms. Destouni received three votes; (2) taking into account both scientific merits and a positive discrimination provision, Ms. Destouni received six votes and Mr Anderson two votes. The selection board proposed that Ms. Destouni be appointed, placing Mr. Anderson in second and Ms. Fogelqvist in third. Later, Ms. Destouni withdrew her application, and the Rector of the University appointed Ms. Fogelqvist to the position. The Rector stated that the difference between Mr. Anderson and Ms. Fogelqvist was not so great as to violate the requirement of objectivity in the selection process. Mr. Anderson and Ms. Abrahamsson brought legal action that eventually came before the Överklagandenämnden för Högskolan, and a preliminary ruling was requested from the CJEU.\footnote{\emph{Abrahamsson}, supra note 13 at § 16-27.}

The Court held that national rules which give a priority to candidates of an underrepresented sex who possess sufficient qualifications for a given post over a candidate of the opposite sex who would have been appointed otherwise on the basis of merit, are precluded under Article 2(1) and (4) of Directive 72/207 and Article 141(4) EC even if the difference between the candidates' qualifications are not so great as to breach the requirement of objectivity. The Court also ruled that national legislation which limits the scope of positive discrimination to a predetermined number of posts, or to posts specifically designed for that purpose, is still precluded because of the absolute and disproportionate nature of the positive discrimination practice.\footnote{\emph{Id.} at § 50-59.}

\section{The (Un)Lawfulness of Fair Automated Hiring}\label{FML}

Fairness metrics are notions of equality mathematically formulated. There exists many fairness metrics, and numerous surveys have been conducted on the pantheon of them  \cite{mehrabi_survey_2021}\cite{pessach_review_2022}\cite{castelnovo_clarification_2022}. Designers of automated hiring systems can implement these metrics during pre-processing or in-processing by curating the data sample or modifying the objective function of the algorithm as was cited in \cite{raghavan_mitigating_2020}, or through post-processing \cite{mehrabi_survey_2021}. If the chosen fairness metric requires the elimination of differences between groups based on a protected attribute while disregarding the base-rate differences between those groups, no matter how the metric is implemented the effect would be to give systematic, preferential treatment to one group at the expense of the other. The frequency or severity of that systematic deviation from equal treatment would be dependent on the strength of the correlation between the sensitive attribute and the target variable in the original, unmodified sample. In fact, the effect of that correlation between sensitive attribute and target variable on the outcome is precisely the ``unfairness'' that must be ``processed'' to achieve the \emph{desired} outcome of group similar results, in spite of group differences.\footnote{\emph{See} \cite{poe_flawed_2023} for a deeper understanding of the underlying analysis.}

The trade-off between an automated hiring system that seeks to achieve equality of treatment versus equality of outcome is inextricably linked to the trade-off between accuracy and fairness that has been widely discussed in the literature \cite{gouic_projection_2020}\cite{zhao_inherent_2022}, where accuracy in employment decisions is a measurable instantiation of qualification assessment objectivity (\emph{Marschall Test}), and fairness in employment decisions is a measurable instantiation of preferential treatment. Placed in this context, the ``cost of fairness'' \cite{menon_cost_2018} is the sacrifice of the individual, fundamental right to equal treatment.\footnote{\emph{See} Charter of Fundamental Rights, Art. 21.} 

Depending on the applicable equality directive and the relevant derogation allowed for by provisions that are in-line with substantive equality of opportunity therein, the lawfulness of the practice that facially violates the principle of equal treatment must be proportional. In the employment context, the CJEU has found that practices (1) which give \emph{absolute} and \emph{unconditional} preference and (2) do not allow for an \emph{objective assessment} of the candidates, ensuring that where candidates are not \emph{equally} or \emph{substantially equally} qualified the preference will be \emph{overridden}, go beyond the limits of the derogation and are thus disproportional.

In \emph{Marschall}, the preferential treatment of the underrepresented sex was limited to tie-breaking scenarios of equally qualified candidates to counteract prejudiced tie-breaking that existed in social reality in accordance with the goal of substantive equality of opportunity. The \emph{Marschall} ``savings clause" ensured that outcome equality was not pursued by requiring the candidatures to be subject to an objective assessment, where the preference would be overridden if the candidate of the over-represented sex had qualifications that would tilt the balance in their favor.

Unlike the practice in \emph{Marschall}, candidates subjected to fair automated hiring processes like the ones described above, are not objectively assessed in the first place, let alone given the assurance of an override. Previous to the selection process and comparison of two applicants, the data sample would have been modified and/or the hypothesis assumptions trained to undervalue the qualifications of some and overvalue the qualifications of others based on their group membership. In other words, the weights of the applicants' features, without preferential treatment, are not brought to bear on the hiring decision. Thus, the integrity of a tie-breaking scenario is compromised at the outset.

In \emph{Abrahamsson}, the CJEU considered legislation which required that a candidate for a public position belonging to the under-represented sex and possessing \emph{sufficient qualifications} for that post must be given a preference over a candidate of the opposite sex that would have been appointed otherwise in order to achieve equal gender representation in the given field of employment. The Court found that the objectivity of the selection process could, therefore, not be precisely determined. Such a practice, the Court reasoned, would result in the selection of candidates with qualifications not equal to but inferior to
those of candidates of the opposite sex, ultimately substituting the individual assessment of candidate merit for group membership. The Court also
ruled that even if the scope of such a practice was limited to a predetermined number of posts, or to posts specifically designed for that
purpose, it would still be precluded because of the absolute and disproportionate nature of the positive discrimination practice.

Unlike in the \emph{Abrahamsson} case, the objectivity, or the lack thereof, of the selection process of an automated hiring system \emph{could} be determined. A system which simply rids the outcome of group skew (defined as the quotient of between group distance and within group distance) or group dissimilarity (in accordance with a given fairness metric) and, in the process of doing so, necessarily disregards the representativness of the sample and robustness of the model, shows its lack of an objective assessment. Further, \emph{Abrahamsson} tells us, that creating a threshold at which candidates are qualified and then ensuring equal outcomes between groups post threshold satisfaction, would likely be precluded under an interpretation of the derogation of the relevant equality directive. In other words, fairness metrics based on statistical sufficiency and separation would likely be just as unlawful as notions based on statistical independence.

In \emph{Kalanke}, the CJEU put forward the primary concern and determining factor of proportionality in the employment context: whether the practice substitutes substantive equality of opportunity for the outcome that is only to be reached by the realization of factual equality in society. For the above reasons, fair automated hiring systems would likely be deemed unlawful due to their \emph{automatic} preferential treatment. Such systems, certainly if used for selection processes of public posts, are simply a high-tech evasion of law which has been settled for decades.  

\section{Conclusion}
As noted in previous research, the dominant underlying definition of ``fairness" in the literature is one of equity and egalitarianism\textemdash equality of outcome rather than either formal or substantive equality of opportunity. Throughout this article, it has been argued that the use of fair machine learning in automated hiring systems for the purpose of providing preferential treatment to some while necessarily positively discriminating against others based on group membership is likely to violate EU non-discrimination law. Meanwhile, as the investigation into commercially available automated hiring systems in \cite{raghavan_mitigating_2020} shows, fair machine learning metrics and techniques are being used by designers of such systems in the present day, potentially resulting in real-world discrimination. Algorithmic fairness and algorithmic non-discrimination are not one in the same, and further research into the conflicts between the two in different jurisdictions and applications is required to ensure that automated decision-making systems are just.

\bibliographystyle{splncs04}
\bibliography{Article}
\end{document}